%% file: ms.tex
\documentclass[journal]{IEEEtran}

\input{packages}

\input{frontmatter}

\author{\IEEEauthorblockN{Lars Larsson\IEEEauthorrefmark{1}, William T{\"a}rneberg\IEEEauthorrefmark{2}, Cristian Klein\IEEEauthorrefmark{1}, Maria Kihl\IEEEauthorrefmark{2}, and Erik Elmroth\IEEEauthorrefmark{1}} \\
\IEEEauthorblockA{\IEEEauthorrefmark{1} Department of Computing Science, Ume{\aa} University, Sweden \\
Email: \{larsson, cklein, elmroth\}@cs.umu.se}
\IEEEauthorblockA{\IEEEauthorrefmark{2} Department of Electrical and Information Technology, Lund University, Sweden \\
Email: \{william.tarneberg, maria.kihl\}@eit.lth.se}}

\begin{document}

\maketitle

\begin{abstract}
\input{sections/abstract}
\end{abstract}


\input{sections/introduction}

\input{sections/stateoftheart}

\input{sections/cachinginfrastructure}

\input{sections/algorithms}

\input{sections/implementation}

\input{sections/evaluation}

\input{sections/relatedwork}

\input{sections/summary}

\bibliographystyle{IEEEtran}
\bibliography{IEEEabrv,references}

\end{document}

%% file: packages.tex
\usepackage{cite}
\usepackage{amsmath}
\usepackage{graphicx}
\usepackage{subcaption}
\usepackage{textcomp}
\usepackage{xcolor}
\usepackage{hyperref}
\usepackage{paralist}
\usepackage{booktabs}
\usepackage{rotating}
\usepackage[figure]{algorithm2e}
\usepackage{siunitx} 

\usepackage[english]{babel}

\usepackage[T1]{fontenc}

%% file: frontmatter.tex
\title{Towards Soft Circuit Breaking in Service Meshes via Application-agnostic Caching}
\author{Lars Larsson \and William Tärneberg \and Cristian Klein \and Erik Elmroth \and Maria Kihl}
\date{}

%% file: sections/abstract.tex
Service meshes factor out code dealing with inter-micro-service communication, such as circuit breaking. Circuit breaking actuation is currently limited to an "on/off" switch, i.e., a tripped circuit breaker will return an application-level error indicating service unavailability to the calling micro-service.
This paper proposes a soft circuit breaker actuator, which returns cached data instead of an error. The overall resilience of a cloud application is improved if constituent micro-services return stale data, instead of no data at all. While caching is widely employed for serving web service traffic, its usage in inter-micro-service communication is lacking. Micro-services responses are highly dynamic, which requires carefully choosing adaptive time-to-life caching algorithms.
We evaluate our approach through two experiments. First, we quantify the trade-off between traffic reduction and data staleness using a purpose-build service, thereby identifying algorithm configurations that keep data staleness at about 3\% or less while reducing network load by up to 30\%. Second, we quantify the network load reduction with the micro-service benchmark by Google Cloud called Hipster Shop. Our approach results in caching of about 80\% of requests.
Results show the feasibility and efficiency of our approach, which encourages implementing caching as a circuit breaking actuator in service meshes.

%% file: sections/introduction.tex
\section{Introduction}
\label{secIntroduction}


Micro-services have emerged as the dominant architectural design pattern for engineering scalable and resilient cloud applications.
Said pattern encourages separation of concerns and data ownership between micro-services~\cite{lewis_microservices_2014}, thus, leading to frequent inter-service requests for data retrieval. In fact, a single public API request may cause orders of magnitude more inter-service requests.

Service meshes~\cite{li2019service} have emerged to factor our commonalities in upstream and downstream communication between micro-services, such as load-balancing, retrying and graceful timeouts. One core feature that is gaining increasing attention is \textbf{circuit breaking}, i.e., reducing downstream traffic in case a condition is detected that suggests overload, such as requests towards a downstream micro-service queueing up.
Circuit breaking consists of three parts: sensors that observe metrics of relevance, a decision mechanism to convert observed metrics in an overload signal, and an actuator that takes action against the overload.

This paper focuses on the \textbf{actuator}. Current circuit breaking actuators are of the "on/off"-type, i.e., a tripped circuit breaker returns a transient application-level error response indicating service unavailability. This paper evaluates an alternative circuit breaker actuator in service meshes: caching. Indeed, cloud application resilience can be improved if constituent micro-services send a reply with stale data, instead of no data at all.

While caching responses is not by itself novel, we are, to our knowledge, the first to suggest and evaluate its usage as circuit breaker actuator in service meshes. Indeed, caching responses is a well-known method for making web content delivery more responsive by reducing service and network load~\cite{chankhunthod1996hierarchical,iyengar1997improving}. However, with the exception of database caching, caching \emph{in general} is not commonly used for inter-service communication.
A key difference between web content caching and inter-service caching is that the latter features highly dynamic responses, that become stale after a few seconds, as opposed to days for web content. Hence, a key question is how the \textbf{time-to-live} (TTL) of a request affects data staleness and network traffic reduction of a realistic cloud application.

To keep the risk of stale data at acceptable levels, we repurpose adaptive TTL estimation algorithms from the web content delivery field for this new purpose (for differences between the fields, please see Section~\ref{secStateOfTheArt}).
To provide a quantitative evaluation of the system, we have selected two dynamic TTL estimation algorithms from the web content caching literature. The selection was driven by plausability to work in the new context of inter-service communication, which has different properties than web content caching (Section~\ref{secAlgorithms}).

To evaluate the two algorithms on a realistic cloud application, we implemented a gRPC-based caching infrastructure, mimicking a service mesh (Section~\ref{secCachingInfrastructure}), and use it to empirically quantify the applicability of caching using dynamically estimated TTLs with two suites of experiments (Section~\ref{secExperiments}). The first suite of experiments (Section~\ref{secEvaluationTrafficReductionDataStaleness}) quantifies the trade-off between network traffic reduction and introduction of errors due to data staleness. To do so, we have developed a service and workload generator that specifically enables data staleness to be controlled and behavior to be observed. 
The second suite applies the conservatively configured algorithms to a realistic micro-service setting (Section~\ref{secExperimentLoadReductionRealistic}). The caching infrastructure is deployed with the Hipster Shop application developed by Google Cloud Platform and use their workload generator to subject the system to simulated e-commerce users.

The contributions of this paper are as follows:

\begin{itemize}
  \item We design and implementation caching as a circuit breaker for service meshes, which works even with gRPC-based communication; as expected from service meshes, the mechanism is application-agnostic and thus requires no source code changes;
  \item We evaluate the inherent trade-offs between network traffic reduction and data staleness using a simple value service.
  \item We demonstrate network traffic reduction with a real micro-service application.
\end{itemize}

The results (Section~\ref{secResultsRealistic}) show that \textbf{about 80\% of inter-service requests could be answered using cached data}, which also caused \textbf{an overall network traffic reduction by 40\%}. Our work suggests that caching is a feasible and efficient circuit breaker actuator, and encourages its implementation in service meshes.

To facilitate reproducibility and reuse of results, we make all our source code and data sets openly available for benefit of the research community. Implementing additional algorithms is a straight-forward process and requires very little code.

%% file: sections/stateoftheart.tex
\section{Background}
\label{secStateOfTheArt}
\label{secBackground}

Caching is extensively used in web content serving, and has been for a long time~\cite{chankhunthod1996hierarchical}. However, it is not commonly used in inter-service communication, and we believe there to be both technical and non-technical reasons for this. The technical ones are temporary hurdles to overcome through engineering: lack of support for caching certain HTTP verbs (gRPC uses POST for every operation, which is typically not considered cacheable), failure to communicate using the right transport protocol (HTTP/1.1 to upstream services rather than HTTP/2), etcetera. All these can be solved rather easily and be incorporated in software. While our work focuses on gRPC, which has no support for caching in its specification (in spite of nascent support in the Protobuf service descriptor for marking operations as idempotent~\cite{grpc_authors_grpc_2019}), it should be noted that there are no \emph{technical reasons} that prevent typical REST-based services from using well-established HTTP/1.1-based caching infrastructure. And yet, it seems to be not commonly done in practice, as exemplified by the fact that even a major vendor like Microsoft does not mention it in their REST API guidelines~\cite{microsoft2020restguidelines}.  

The non-technical reasons are more interesting to us, as the major hurdle does not seem to be the technical challenges. A reason that cannot be ignored is that \textbf{it is hard to \emph{a priori} determine TTLs for responses}. Software developers cannot during development reasonably know for how long responses will be valid, unless the underlying data is known to be stable for some time (e.g.\ weather estimates that are updated hourly). But letting software inspect responses and thereafter estimate TTLs during runtime is definitely possible, as our results show.

Determining \emph{which} operations are possible to cache can also present a challenge. It is generally considered good API design to separate operations that can mutate state from the ones that cannot. REST enforces this via HTTP verb mapping~\cite{fielding2000architectural}. Because gRPC lacks caching on the protocol level, there is no such enforcement. Still, it is an ingrained best practice design pattern and developers and operators are therefore generally aware of which operations can mutate state and can therefore inform software of it. 

It is a generally accepted practice to use a fast in-memory key-value store such as Redis in front of databases for read queries to avoid needlessly straining the database service with possibly complex queries (e.g., ones requiring multi-table JOINs)~\cite{familiar_iot_2015,brown_implementation_2016}. The application code is then adapted to always check the key-value store \emph{before} issuing the possibly complicated database query, where the results may be cached. Thus, it is up to application developers to not only decide which operations to cache but also, possibly, for how long. The approach we take differs in that we
\begin{inparaenum}[(a)]
\item cache in-between services, not just in front of the canonical database server; 
\item require no application awareness of caching --- as, indeed, gRPC applications have no concept of caching; and
\item object cache time-to-live is continuously re-estimated.
\end{inparaenum}
In this way, applications can get the benefits of caching across micro-service architectures where calls are performed in many steps before hitting a database, and application developers need not make their applications cache-aware.

Services that use gRPC for inter-service communication often expose a REST interface toward clients. Would it therefore not be sufficient to cache only the client-facing responses? We argue that it is \textbf{not sufficient} in a micro-service application, for two reasons. Firstly, modern services typically have analytics and other batch jobs that rely on direct inter-service requests, rather than on publicly facing aggregated APIs. Second, TTLs for aggregated results are bounded by the lowest TTL among the constituent sub-results. Our results with the Hipster Shop application show that, on average, a single client request branches out and requires aggregated data from around 13 inter-service requests (Section~\ref{secResultsRealisticTrafficAnalysis}). Should even a single of these have low TTL and the others a high TTL, it would invalidate the aggregated response and all requests would have to be wastefully re-issued if only client-facing caching was used. This has been previously explored with regard to personalized web sites in e.g.~\cite{shi2002workload}.

Because inter-service communication differs from web content caching, we regard the following properties as important differences:

\begin{itemize}
  \item updates are potentially more frequent, and TTLs therefore shorter.  Well-designed web applications consist of immutable and therefore infinitely cacheable static resources, presented via a dynamic HTML page, making the orders of magnitude smaller (in bytes) HTML page the only asset that needs a short TTL;
  \item large variance in object popularity. Unlike web content caching, where some objects are much more popular than others due to human preference (70\% of objects at a CDN were requested only once over a multi-day period~\cite{basu2018adaptive}), API requests are highly varied and popularity distribution need not be tied to human preference; and
  \item calculations and responses must be fast. Because client-facing requests cause multiple inter-service requests, caching must not add significant delays, lest the multiplicative effect be noticeable. 
\end{itemize}

%% file: sections/cachinginfrastructure.tex
\section{Soft Circuit Breaker Actuator: Architecture}
\label{secCachingInfrastructure}
\label{secSystemArchitecture}

\input{figures/architectureOverview}


Service meshes instantiate for each micro-service two proxies: one handling upstream calls and one handling downstream calls. This allows the service mesh to intercept all inter-service calls in an application-agnostic way, and offer higher-level communication functionality, such as circuit breaking. We hereby propose a system architecture that can readily be deployed in a service mesh.

Our proposed caching circuit breaker actuator has two components: the \emph{Estimator} and the \emph{Cache}. Respectively, they are responsible for estimating for how long a response object is valid and for caching responses for (maximally) that amount of time.

Figure~\ref{figArchitectureOverview} shows the conceptual architecture and traffic flow in the system. Instead of direct connection between the Client and Server (marked with dotted gray line), the newly added components are deployed and configured to intercept the traffic (black lines).

The Estimator and Cache components can be deployed in different configurations, i.e., in the downstream or upstream proxy of the service mesh, each favoring different aspects of a performance and cache coherency trade-off. These are discussed in Section~\ref{secDeployment}.

\subsection{Cache component}
\label{secCacheComponent}

Unlike HTTP/1.1, where caching is specified as part of the protocol~\cite{rfc2616}, gRPC has no notion of caching (see also Section~\ref{secStateOfTheArt}). Accordingly, the Cache component in our proposed system must respond as the Client expects a Server to respond. This makes the Cache behave indistinguishably from a Server from the point of view of the Client, thereby allowing for seamless integration with existing gRPC applications.

It is valid to add metadata in headers for gRPC responses. We use the \texttt{Cache-Control} header to express the TTL in seconds, similar to how HTTP/1.1 defines it. If response TTL is given in the header of a response, the Cache component will cache the response for the given amount of time. If not, or the TTL is specified as 0, the response will not be cached. 

\subsection{Estimator component}
\label{secEstimatorComponent}

The Estimator component estimates how long a response to a particular request can be considered valid and therefore cached. Since gRPC Servers do not typically convey how long responses are valid, the Estimator can use multiple different algorithms to estimate object cache validity (see Section~\ref{secAlgorithms}).

When a request has been made to the Estimator, it will for a limited duration of time produce response TTL estimates for subsequent equivalent requests. The time limit is used for housekeeping purposes: once the time limit is surpassed, the Estimator will de-allocate the memory used to calculate estimates for the particular request.

Because the Estimator cannot know when a response to a request has changed, it has to continuously update its estimates. The Estimator will contact the upstream Server whenever it gets an incoming request. The reason for an incoming request must be that the Cache cannot answer a Client request from memory, which either means that the Cache has restarted or the response TTL has been surpassed. Regardless, the Estimator will contact the upstream Server and make a new TTL estimate. 

\subsection{Component co-deployment}
\label{secDeployment}


Because the Cache and Estimator components are designed to seamlessly deploy into the network between Client and Server, a number of different deployment scenarios are possible. In this work, we focus solely on the case where Cache components are \emph{co-deployed} with Clients, and Estimator components with Servers. We defer investigation into the consequences of the different deployment scenarios with regard to, e.g., cache consistency and traffic reduction to future work. In practical terms and in the context of this work, co-deployment means that a \emph{sidecar} container is started in a Kubernetes Pod. By definition, this implies that localhost networking can be used between co-deployed components. This follows an established pattern of how, e.g., service meshes such as Istio offer their services.

%% file: figures/architectureOverview.tex
\begin{figure}
\centering
\includegraphics[width=0.6\columnwidth]{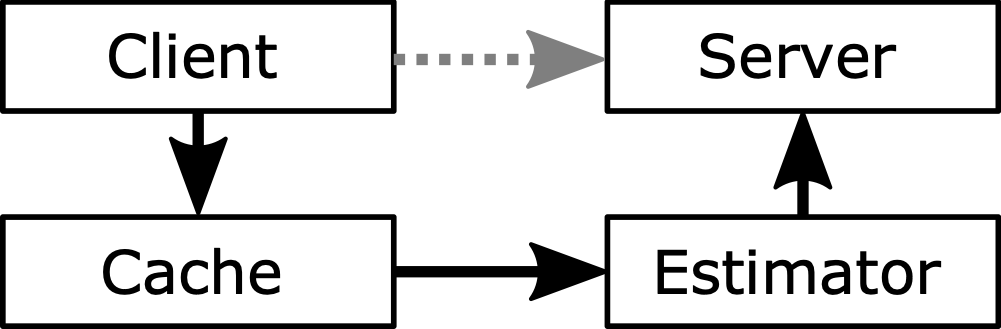}
\caption{Caching infrastructure architecture overview showing the old traffic flow as a dotted gray line and the new traffic flow through Cache and Estimator components in black.}
\label{figArchitectureOverview}
\end{figure}

%% file: sections/algorithms.tex
\section{TTL estimation algorithms}
\label{secAlgorithms}

Meeting the requirements stated in the previous section and cognizant of differences between web content serving and inter-service request handling, we have implemented to algorithms that take very little memory and require no large body of training data to function. For comparison reasons, we have also implemented a simple static TTL ``estimation'' as well.



\subsection{Static TTL}
\label{secAlgorithmsStatic}

The Static TTL algorithm acts as a point of reference and base case. It takes as runtime configuration a single parameter, $\beta$, namely the number of seconds (integer) to always statically respond with as response TTL for each incoming request. Thus, the TTL of an object $x$ ($\textrm{TTL}_{x}$) simply is:

\begin{equation}
\textrm{TTL}_{x} = \beta
\label{eqStaticTTL}
\end{equation}

Note that setting the $\beta$ parameter to zero implies that no caching should be made. In doing so, we obtain a base case where the caching infrastructure is in place, but no caching is performed. Thus, neither network traffic reduction nor data staleness are introduced. 

\subsection{Adaptive TTL}
\label{secAlgorithmsAdaptive}

The Adaptive TTL algorithm~\cite{lee2002updaterisk} uses a simple heuristic to estimate TTL of an object $x$ ($\textrm{TTL}_{x}$), based on the time interval between when the object was last modified ($M_{x}$) and the current time ($t$). It is parameterized by $\alpha$, a real number that, while technically semantic-free~\cite{lee2002updaterisk}, practically signifies a linear ``acceptance'' of stale data by the operator. Higher values of $\alpha$ mean longer estimated TTL, and thus, higher risk of stale data. The estimated TTL is given by Equation~\ref{eqAdaptiveTTL}:

\begin{equation}
\textrm{TTL}_{x} = (t - M_{x}) \times \alpha
\label{eqAdaptiveTTL}
\end{equation}

While per definition $\alpha$ can take on any positive real value, it is unlikely to be useful if larger than 0.5. To see why, consider that $\alpha = 0.5$ states that if the last modification was 10 second ago, 5 seconds would be a reasonable TTL. This causes the system to behave in a manner inspired by the Nyqvist-Shannon Sampling Theory~\cite{shannon_communication_1949}: sampling twice as often as (is estimated to be) needed. 

We have implemented the Adaptive TTL algorithm as described in~\cite{lee2002updaterisk}. The memory requirements per request of the implementation includes only storing a hash value of the latest updated response and the associated time stamp when the response last changed. 

\subsection{Update-risk based TTL}
\label{secAlgorithmsUpdaterisk}

Lee et al.\ introduced an Update-risk based TTL estimation scheme in~\cite{lee2002updaterisk}. According their paper, choosing a good value of $\alpha$ in Adaptive TTL requires guesswork and suffers because there is no clear semantic meaning to $\alpha$. Instead, the Update-risk based algorithm takes as a parameter an operator-specified \emph{acceptable update risk}, $\rho_{x} \in [0,1)$, for a given object $x$. Low values result in low TTLs and therefore low risk of missing an update (stale data). Values close to 1 implies that a large update risk is allowed, resulting in a very large TTL, and consequently significantly higher risk of data staleness.

If we let $\textrm{BUD}_{x}(K)$ signify the ``backward K-update distance'' (point in time of the $K^\textrm{th}$ most recent response object update of $x$), then the estimated TTL for object $x$ ($\textrm{TTL}_{x}$) is given by Equation~\ref{eqUpdateRisk}:

\begin{equation}
\textrm{TTL}_{x} = - \frac{\textrm{BUD}_{x}(K)}{K}\log(1-\rho_{x})
\label{eqUpdateRisk}
\end{equation}

By experimentation, the original authors found that $K=2$ provides the best estimates~\cite{lee2002updaterisk}, and that is what we use in our implementation as well. Intuitively, this means that the calculation determines rate of change by keeping a history of two modification timestamps and then dividing by 2. The implementation therefore only requires keeping two response hashes and associated timestamps per request.

The Update-risk based one is \emph{less reactive} than Adaptive TTL in estimating update frequency, which only uses the timestamp of the single most recent modification to do the same. It should also be noted that Lee et al.\ mathematically prove that setting $\rho = 1 - e^{-\alpha}$ and using $K=1$ makes Update-risk based behave as Adaptive TTL~\cite{lee2002updaterisk}. We did however not verify that via practical implementation, and rather implemented them separately.

%% file: sections/implementation.tex
\section{Implementation}
\label{secImplementation}

Our design goals for the implementation are to be extensible, suitable for research via instrumentation/observability, and easy to integrate with existing service meshes. The latter implies an application-agnostic approach, such that existing gRPC-based services can benefit from it without source code modifications.

\emph{Extensibility} is ensured via implementing the Estimator and Cache gRPC \emph{interceptors}, i.e., as plugins that capture and possibly modify requests before they are passed along to the intended process. Interceptors can be chained, thus allowing other interceptors to also impact requests and responses, which may be required to, e.g., maintain information enabling distributed tracing.

\emph{Instrumentation/observability} for research is implemented by letting interceptor output timestamped CSV rows with nanosecond resolution. All operations output the name of the invoked method. The Cache also states whether a response had to be passed upstream to the Estimator or could be answered using cached data. In addition to method name and timestamp, the Estimator outputs the TTL estimate for a given response. Together, the data can be used to form a picture of overall system performance. 

\emph{Application-agnosticism} and the ability to use our caching infrastructure without source code modification is enabled by attaching the Cache and Estimator gRPC interceptors to purpose-built reverse proxies. The code for these is auto-generated from the Protobuf service descriptor using our modified version of the gRPC code stub generator. This way, observability and configurability is also increased, because message contents can be fully inspected and used by the interceptors. This enables processing such as, e.g., blacklisting operations from caching based on the presence of some named attribute such as ``user\_id''.

\subsection{Cache protocol}
\label{secCacheProtocol}

As previously mentioned, the Cache must serve responses to the Client just as a Server would have done, because gRPC applications are unaware of caching. Therefore, the Cache always serves the full response object when requested along with an \texttt{HTTP 200 OK} response code. 
For possible compatibility with third-party systems (see Section~\ref{secStateOfTheArt}), current or future, the Estimator communicates with the Cache via the \texttt{Cache-Control} header introduced in HTTP/1.1~\cite{rfc2616} and which is valid also for HTTP/2. As such, object cache lifetimes are expressed in integer values of seconds, according to its specification.  
The Estimator does not honor \texttt{Max-Age} headers attached to requests, because whenever it receives a request, it will \emph{always} forward it to the upstream Server and respond with the freshest possible data. This is intentional, as it is the mechanism by which the Estimator can learn about response updates and make new TTL estimates.
Because query requests that do not modify application state are the only ones that can safely be cached, we offer an optional blacklisting functionality in the Estimator so that operations can be excluded from caching entirely. 


\subsection{Limitations}
\label{secImplementationLimitations}

Our implementation is a proof of concept for research purposes, and as such, contains some simplifications. Simplifications inevitably introduce limitations, some of which are discussed below.

Although existing gRPC Clients would not be able to use the \texttt{If-Modified-Since} flow HTTP/1.1 introduced~\cite{rfc2616}, nothing would prevent our Cache component from using it toward the Estimator. However, our current implementation does not support it. Instead, the Estimator will always, if called, fetch a new response from the Server. The Estimator assumes that the Cache will only call it if the Cache \emph{needs} to get a fresh copy, e.g., if the object has expired or the Cache has restarted. To be compatible with current or future general-purpose HTTP/2-enabled caching services, we leverage the \texttt{Cache-Control} header as specified in the HTTP/1.1 specification~\cite{rfc7234} (which HTTP/2 uses) to communicate TTL. Thus, we are limited to integer values of seconds for how long a response can be cached.

Our current implementation targets unary gRPC operations, not streaming ones. The latter is conceptually similar to the former, but handled differently on a technical level only. To the best of our knowledge and/or imagination, none of the limitations listed here should impact the validity of our results. Caching for fractions of a second would absolutely produce \emph{different} results, but the \emph{validity} of the results, given these limitations, is not impacted.

%% file: sections/evaluation.tex
\section{Evaluation}
\label{secEvaluation}
\label{secExperiments}

The objective of this section is to establish a set of experiments that will validate the proposed caching infrastructure concept and its implementation.
Further, because caching always introduces a risk of stale data to achieve a reduction in network traffic, the inherent trade-offs in our proposed system must be evaluated and addressed. In particular, we must establish which trade-offs are provided by which algorithm configurations and whether the dynamic caching and supporting infrastructure approach work for real micro-service applications. 

First, we quantify the trade-off between network traffic reduction and data staleness. The first suite of experiments are designed to legitimize the proposed caching infrastructure and give guidance to how to configure and tune the TTL estimation algorithms. To do so, we designed a bespoke service and workload generator. Using these, we are able to eliminate noise and uncertainty inherent in large deployments and have full observability. For these experiments, we compare the dynamic behavior of the algorithms to three static baselines, as well as no caching.

Second, to validate the caching infrastructure in a real setting, we perform experiments with a real micro-service application. We have chosen the ``Hipster Store'' by Google Cloud Platform. The focus of this experiment is two-fold:
\begin{inparaenum}[(a)]
\item to verify that our caching infrastructure works for a micro-service application without any source code modification; and
\item to see what network traffic reductions can be made using caching and conservatively configured dynamic TTL estimation algorithms.
\end{inparaenum}

Data staleness, while often favorable compared to non-responsive services, is generally to be avoided. However, what level of staleness is acceptable is application-dependent: certain values are never allowed to be stale (e.g.\ a customer's order history), whereas others are less critical (e.g.\ a product recommendation). In the first suite of experiments, \emph{any} data staleness is regarded as an error, as quantifying data staleness vs.\ network traffic reduction is what the experiment is clearly designed for. In the second suite, however, intended application-specific data staleness sensitivity of the various micro-services is not known to us. We therefore do not quantify data staleness as part of the second suite of experiments, but rather, conduct the experiments using only conservatively configured TTL estimation algorithms informed by the first suite of experiments, to keep staleness as generally low as possible. 

For general applicability, we do not explicitly focus on latency or response times as part of our analysis. Latency and response times are nonlinear functions of the amount of work that a server has to do~\cite{1191656} and depend on a multitude of factors, such as application code, its deployment, and the underlying hardware resources, the confluence of which causes unexpected behavior in both the application and the control plane~\cite{larsson_impact_2020}. Thus, a more objective and general measurement on algorithm efficiency and performance is to consider the number of requests that are transmitted across the network and, when a specific application is used, the number of bytes such transmissions consist of, rather than focusing unduly on the time these transmissions and request processing takes. Unless, of course, caching \emph{itself} would add considerable processing time --- however, our choice of algorithms (Section~\ref{secAlgorithms}) and results (Section~\ref{secResultsRealisticTrafficAnalysis}) strongly indicate that this is not the case here.

\subsection{Quantifying trade-offs between network traffic reduction and data staleness}
\label{secEvaluationTrafficReductionDataStaleness}

\input{figures/sinusoidal-workloads}

In the first suite of experiments, we deploy a Client, Server, and interconnecting caching infrastructure (Cache and Estimator) like in Figure~\ref{figArchitectureOverview}. The Client and Server constitute a very simple custom-made service that keeps track of a single value, the \emph{Value Service}. Its simplicity allows us to accurately measure and control data staleness. 

The Server of the Value Service exposes a request method (\texttt{GetValue()}) and an update method (\texttt{SetValue(newValue)}) and is multi-threaded. The Client of the Value Service uses two independent threads. One for querying the value from the Server, and one for updating it. 

The two Client threads share a value in a concurrency-safe way, such that the query thread can safely read what the true value should be as dictated by the updater thread. The query thread can therefore determine whether a response matched the expected value. This lets us calculate the error fraction using Equation~\ref{eqErrorFraction}:

\begin{equation}
  \textrm{error fraction} = \frac{\textrm{\# queries with non-expected value}}{\textrm{\# total queries}}
\label{eqErrorFraction}
\end{equation}

To measure network traffic reduction, we assume that the Client and Cache are co-deployed (see Section~\ref{secDeployment}) in one machine or Kubernetes Pod, and that the Estimator and Server are co-deployed in another. Thus, we calculate network traffic reduction as in Equation~\ref{eqNetworkTrafficReduction}: 

\begin{equation}
  \textrm{network traffic reduction} = \frac{\textrm{\# queries answered by Cache}}{\textrm{\# total queries}}
\label{eqNetworkTrafficReduction}
\end{equation}

\subsubsection{Workload generation and repeatability}
\label{secExperimentsWorkloadGeneration}
\label{secExperimentsRepeatability}

The workload consists of two types of Client requests: updates or queries. The rates at which it does are referred to as the \emph{update rate} and the \emph{query rate}, respectively. Two properties of the workload are important with regard to caching:
\begin{inparaenum}[(a)]
\item the relationship (ratio) between update and query rate; and
\item how the rates change over time.
\end{inparaenum}

The inter-rate relationship matters because the \emph{potential benefit} of caching is dependent on a high query rate (more requests answered from cache), but the \emph{possibility} to cache is dependent on a low update rate (to avoid data staleness). The overall utility depends on the ratio between the rates, as lowering the query rate reduces the potential benefit regardless of update rate, and increasing the update rate reduces the possibility to cache regardless of query rate. 

However, conducting an evaluation merely studying a selection of inter-rate ratios would be both unfair and misleading: not only do real users not behave like that, but some of the algorithms will be at a clear advantage in that case. Consider Equation~\ref{eqAdaptiveTTL} describing the Adaptive TTL algorithm. When an update is detected, it immediately reacts by lowering TTL estimates to 0. Thus, it will be called again in the immediate future, and be unlikely to ever miss an update, resulting in neither network traffic reduction nor data staleness --- given that behavior would not change over time, the length of the experiment would then not matter, except to artificially inflate the resulting numbers.

To both avoid unfairness and misleading results and to capture the fact that web workloads are non-trivial and dynamically change over time, a caching system must be able to deal with update and query rate changes over time. To compactly study this behavior, we use a sinusoidal workload pattern. 
Both query and update threads in the Client use an individually parameterized sinusoidal function to calculate their respective rates.
Our experiments vary the mean update rate between \SIrange{0.05}{1.1}{} (updates per second), and the mean query rate between \SIrange{1}{10}{} (queries per second). This constitutes a large span of values, with a factor 200 difference at its highest (0.05 updates vs.\ 10 queries per second). 
Additionally, we are interested in the performance of the TTL estimation over all possible scenarios, including update rates are low and queries are high and vice versa. 
Therefore, we also shift the phase of the sinusoidal functions to offset the peaks and valleys in query and update rates.

The experiments are each 30 minutes long (1800 seconds), constituting a full period for the aforementioned sinusoidal workload.
The duration was chosen because the time-scale and update frequency is significantly shorter than for, e.g., web content caching, where durations of days are more common in experiments (Section~\ref{secRelatedWork}).
Figure~\ref{figSinusoidalWorkloads} shows mean update and query rates for the duration of the experiments. 
We phase shift the mean update rate from full to no alignment, namely by $0, \pi, \pi/2,$ and $\pi/4$ radians. 
Intuitively, a full alignment (0 radian phase shift) means that query and update rates are both high and low at the same time. 
This is a service that experiences a given ratio of updates and queries at all times. 
In contrast, at a phase shift of $\pi$, the update rate is high when the query rate is low and vice versa. 
This represents the largest possible variation in difference between update and query rates --- at times there are even more updates than there are queries. Thus, although the range of values is large, our experiments subject the system to different symptomatic types of workloads, e.g., read- or write-heavy, high or low overall intensity, etc.

For each request, we sample the integer request delay using a Poisson process drawn from the sinusoid at that time instant, shown in Figure~\ref{figSinusoidalWorkloads}. Relying on the Poisson distribution here both mimics the true behavior of client-server systems more closely than a sinusoidal function does, and makes our results comparable to other literature in the network performance and queuing theory fields~\cite{harchol2013performance}. 

The following parameters uniquely describe an experiment quantifying the trade-off between network traffic reduction and data staleness:
\begin{inparaenum}[(a)]
  \item the algorithm used;
  \item configuration of said algorithm; and
  \item phase shift of update thread in relation to the query thread.
\end{inparaenum}

We repeat each experiment 3 times with different deterministically set seed values used for drawing from the Poisson distributions determining true update and query rates. We then take the mean from these 3 repetitions with different seed values as the results for each experiment. The full data set and all scripts required to re-run the experiment in full are available as an open dataset.

\subsubsection{Algorithm configurations}
\label{secExperimentsConfigurations}

The following algorithm and algorithm parameter combinations were used during the experiments on quantifying the trade-off between network traffic reduction and data staleness:

\begin{itemize}
  \item \textbf{static-0}, which is the cache-free base case, where the static TTL parameter is set to 0.
  \item \textbf{static-1}, \textbf{static-10}, and \textbf{static-30}, setting the static cache time parameter to 1, 10, and 30 seconds, respectively. The latter two are intended to show upper bounds on caching-related behavior. It is worth noting that the lowest possible update rate in our experiments is 0.05 updates per second, i.e.\ updates that are 20 seconds apart. Caching for 10 seconds would, in these extreme cases, be a very good choice (half the update rate, inspired by the Nyqvist-Shannon Sampling Theorem~\cite{shannon_communication_1949}). Finally, caching for 30 seconds will always be bad from a data staleness perspective, but reduce network traffic significantly.
  \item \textbf{adaptive-0.1}, \textbf{adaptive-0.25}, and \textbf{adaptive-0.5}, which is the Adaptive TTL with $\alpha$ set to 0.1 (a limit mentioned in~\cite{lee2002updaterisk} and incorrectly attributed to the HTTP/1.1 specification), 0.25 as a conservative midway point, and 0.5, after findings by~\cite{chankhunthod1996hierarchical}. Note that Adaptive TTL uses its parameter linearly in estimating TTL (see Section~\ref{secAlgorithmsAdaptive}).
  \item \textbf{updaterisk-0.1}, \textbf{updaterisk-0.25}, \textbf{updaterisk-0.5}, \textbf{updaterisk-0.75}, and \textbf{updaterisk-0.90}. The relatively larger number of configuration parameters is due to the non-linear behavior of the Update-risk based algorithm with regard to its $\rho$ parameter (see Section~\ref{secAlgorithmsUpdaterisk}), which is harder to reason about without having access to the underlying experimental data.
\end{itemize}

\subsubsection{Results: trade-off between network traffic reduction and data staleness}
\label{secResultsNetworkReductionDataStaleness}

\input{figures/network_reduction_vs_error_fraction}
The results of the experiments are presented below, followed by an analysis in Section~\ref{secDiscussionNetworkReductionDataStaleness}.

Figure~\ref{figNetworkReductionVsErrorFraction} shows the results of the experiments that evaluate the trade-off between network traffic reduction and error fraction. Again, for observability, here we use our bespoke Value Service. The size of the markers correlate with the value of the algorithm's configuration parameter. All phase shifts are included, which is why for each algorithm and size of marker, there are four such markers.

With regard first to \textbf{network traffic reduction}, Figure~\ref{figNetworkReductionVsErrorFraction} shows that by merely issuing 1 second TTL for all responses, an 85\% reduction is achieved. Because the static algorithms do not dynamically calculate TTL based on, e.g., update rate, the reductions for a given parametrization are the same across all Client updater thread phase shifts.

The dynamic algorithms, Adaptive TTL and Update-risk based, denoted in figures as ``dynamic-adaptive'' and ``dynamic-updaterisk'', respectively, show a wider range of network request reduction. In all parametrizations, the case when updater and query threads move in synchrony (see Figure~\ref{figSinusoidalWorkloads} with phase shift 0 radians) is the one that allows for the least amount of traffic reduction. Conversely, phase shift $\pi$, the case where update rates are at their highest and query rate is at its lowest and vice versa, shows the largest amount of network request reduction. Across the most conservative configurations of the respective dynamic algorithms (their parameters set to 0.1), traffic reduction is between 3--30\%, with an average of 17\%. 

The \textbf{error fraction} dimension of Figure~\ref{figNetworkReductionVsErrorFraction} shows that by not taking update rate into regard and always assigning a 1-second TTL (\textbf{static-1}), the error fraction is between 12--26\% in these experiments. The figure also shows that the less conservative algorithm parametrizations, which produce larger TTL values of many seconds, are associated with a much higher incidence of data staleness. Conservatively configured dynamic algorithms can, on the other hand, keep error fractions in the low, single-digit, percentages. Analysis of the data shows that the reason is that they scale down TTL estimates to zero when the update rate is too high. This might seem counterintuitive, but is in-fact a reflection of the nature of the algorithm. The greater the $\rho$ the larger the risk. 

\subsubsection{Analysis: trade-off between network traffic reduction and data staleness}
\label{secDiscussionNetworkReductionDataStaleness}

The results presented in the previous section show that caching using dynamically estimated TTLs can greatly reduce the number of requests that must be transmitted over the network, while also keeping data staleness low. Considering \emph{only} the network reduction in Figure~\ref{figNetworkReductionVsErrorFraction}, the \textbf{static-1} algorithm showed great promise in its simplicity. Simply always caching for 1 second reduced the number of queries during the experiments by 85\%! However, this trivial algorithm \emph{also} introduces an error fraction of upwards of prohibitive 25\%! In contrast, the dynamic TTL estimation algorithms manage much better with regard to data staleness. 

\input{figures/dynamic-updaterisk-phaseshift-pi.tex}

But what causes the dynamic algorithms to keep error fraction low but still makes them able to reduce network traffic? Choosing two experiments with the same phase shift and algorithm, but with different parameters, highlights this.

Figure~\ref{figUpdateRiskTimeseries} shows the behavior of the Update-risk based algorithm in more detail for two of its parametrizations and for the $\pi$ phase shift. The more conservative configuration ($\rho=0.1$) estimates TTLs with 1-2 seconds at most but most often zero, whereas the configuration that accepts higher risk ($\rho=0.5$) increases TTL estimation significantly. While that achieves a much higher traffic reduction, it also pays the price in data staleness, with an unreasonably high error fraction throughout the experiment, except for when the cache-friendly circumstances of a peak in query rate and a valley in update rate occur at around 300 seconds into the experiment. Thus, given a sufficiently high query rate, any caching, even if just a single second, will yield great network traffic reductions.

In summary, both dynamic TTL estimation algorithms, when configured with low parameter values, keep data staleness errors at single-digit percentages while also reducing network traffic by upward of 30\%.

\subsection{Quantifying network load reduction in a real application}
\label{secHipsterShopExperiments}
\label{secExperimentLoadReductionRealistic}

Although it would be desirable to evaluate data staleness in this experiment too, this is difficult to achieve. For example, if the client pays for an item to the shopping cart, this not only affects the Cart micro-service, but also a number of others, e.g., the ShippingService (see Figure~\ref{figHipsterShopOverview}). Hence, even if we modified the client to remember the last value set for each API call, side-effects across micro-services prevent the client from accurately predicting the freshest value for a different API call. Therefore, we shall use the insight on data staleness we gathered from the first experiment.

Based on the findings in the first suite of experiments, our conclusion is that the more conservative parametrizations should be the most \emph{generally applicable} in practical settings, bearing in mind that sensitivity to stale data is application-specific (Section~\ref{secEvaluation}).

The second suite of experiments aim to show the correct functioning of our caching infrastructure with a real micro-service application and to quantify the network reduction benefits that may be possible when caching with conservatively estimated TTLs are used. 

Hipster Shop, chosen for our evaluation, is a polyglot application that consists of 11 micro-services that communicate over gRPC. See Figure~\ref{figHipsterShopOverview} for an architectural overview. 

Note that \textbf{no source code has been modified} in Hipster Shop. We only modified the Kubernetes deployment manifest such that our dynamic caching infrastructure is put in place and services communicate through it. These modifications are simple and can be automated in the future, e.g., via a Kubernetes Mutating Admission Controller like contemporary service meshes do.

\input{figures/hipsterShopOverview}

To each service in Hipster Shop, we added an Estimator. Because not all requests can be cached, we blacklisted certain requests (see also Section~\ref{secHipsterShopCacheable}). The Estimator components were all given the same configuration, depending on which algorithm was under test (see Section~\ref{secHipsterShopConfigurations}).

Cache components were added to the three services in Hipster Shop that perform inter-service calls over gRPC (marked with bolder borders in Figure~\ref{figHipsterShopOverview}): Frontend, CheckoutService, and RecommendationService. Adding a Cache component to other services would not affect them in any way, apart from wasting resources on a component that would be dormant. Note that we \textbf{do not cache non-gRPC traffic}, i.e.\ the HTTP responses to the Load Generator and the Redis communication that the CartService engages in with its database. 

\subsubsection{Algorithm configurations}
\label{secHipsterShopConfigurations}

Results from the experiments using the Value Service showed that only the most conservative configurations keep data staleness relatively low. As shown in Section~\ref{secResultsNetworkReductionDataStaleness}, statically caching responses for even a single second introduces data staleness upward of 25\%, which we deem unacceptably high for general applications.

Therefore, the algorithm configurations used for this experiment were the two configurations of dynamic TTL estimation algorithms that introduced the least amount of data staleness errors, namely Adaptive TTL with $\alpha=0.1$ and Update-risk based with $\rho=0.1$. For reference, we also deployed the system with the caching infrastructure in place and caching disabled (statically set TTL to 0).

\subsubsection{Load generation}
\label{secHipsterShopLoadGeneration}

Hipster Shop ships with its own load generator. The load generator operates in closed-loop manner~\cite{schroeder2006open} and waits a random amount of time (uniform distribution) before issuing the next request. The set of possible requests is pre-defined, and weights are attached to the requests, which affects the probability that a particular request is randomly chosen more or less often than the others.

Because the aim of this experiment is to show caching infrastructure compatibility and potential network traffic reduction, we used the Hipster Shop load generator as-is. This way, we neither introduce errors due to misleading assumptions about application or user behavior nor skew results in any particular way. We configured it to simulate 100 concurrent users, rather than the 10 users that it is set to by default, because we did not merely want a trickle of background traffic but a workload that resembles a modestly popular boutique e-commerce site.

\subsubsection{Cacheable subset of operations}
\label{secHipsterShopCacheable}

Like in the Value Service, not all operations in Hipster Shop can be cached without introducing significant application-level errors. Caching, e.g., a call to \texttt{AddItem(user\_id, item)} such that a repeated call would be ignored by the CartService would be highly detrimental to the application: the client would get a response indicating success, but the CartService would not have registered the intent to put the item in the cart of the given user.

To mitigate this, we used the cache blacklisting feature described in Section~\ref{secImplementation} to disallow caching of such state-modifying calls. Continuing with the CartService example, \texttt{GetCart(user\_id)} would be possible to cache, but \texttt{AddItem(\ldots)} and \texttt{EmptyCart(\ldots)} would not be. Due to caching, calls to \texttt{GetCart(user\_id)} might return the incorrect response due to data staleness, but will eventually be corrected because the two state-modifying operations were passed to the CartService. For the full detailed list of operations that we blacklisted, we refer readers to the source code repository holding our experimental setup.

\subsubsection{Results: network traffic reduction for a real micro-service application}
\label{secResultsRealistic}

\input{figures/hipsterShopNetworkTraffic}

The results of the experiments are presented below, followed by an analysis in Section~\ref{secResultsRealisticTrafficAnalysis}.
Based on the results obtained by the previous sets of experiments, we deployed Hipster Shop in a Kubernetes cluster (minikube instance) with our caching infrastructure. The two most conservative dynamic TTL estimation algorithm configurations were deployed (``dynamic-adaptive-0.1'' and ``dynamic-updaterisk-0.1'') as well as the no-caching baseline (``static-0'').

\input{tables/hipsterShopNetworkTraffic}

Figure~\ref{figHipsterShopNetworkTraffic} shows the amount of network traffic for the three algorithm configurations, averaged in 15-second increments during the three experiment repetitions. The amount of network traffic is visibly clearly reduced when caching is used, in comparison to when it is not. Table~\ref{tabHipsterShopNetworkTraffic} shows key values regarding network traffic, also averaged over the three experiment repetitions. Both caching algorithms achieve a traffic reduction of about 40\%, with the very slight advantage going to the Adaptive TTL algorithm.

\input{tables/hipsterShopCachedRequests}

What caused the traffic reduction is the use of cached data. Table~\ref{tabHipsterShopCachedRequests} shows caching of requests in our experiments. The total number of requests in the application are very similar across experiments (differences related to randomness in the load generator, see Section~\ref{secHipsterShopLoadGeneration}) and both dynamic algorithms manage to cache about 80\% of responses. As stated in Section~\ref{secHipsterShopCacheable}, not all requests can be cached. Service responses are of course also not equal in size, which explains why an 80\% reduction in requests could translate into a 40\% reduction in network traffic.

\subsubsection{Traffic analysis}
\label{secResultsRealisticTrafficAnalysis}

Analysis of the Hipster Shop experiment log files show that 12\% of requests were not initiated by the Frontend service. Recalling Figure~\ref{figHipsterShopOverview}, two other services (RecommendationService and CheckoutService) also make requests to carry out their work. This implies that inter-service requests that can be answered from cache between these services and the ones they request data from benefit from not only caching at the publicly facing Frontend.

Although we purposefully did not seek out to include response time analysis in these experiments (for good reason: the minikube Kubernetes cluster is far from a production-ready or realistic execution environment), it is worth noting that response time for most operations was cut in half with caching enabled compared to when it was not (not shown for briefness and to not place undue focus on it in this evaluation).

%% file: figures/sinusoidal-workloads.tex
\begin{figure*}
  \centering
  \includegraphics[width=0.85\textwidth]{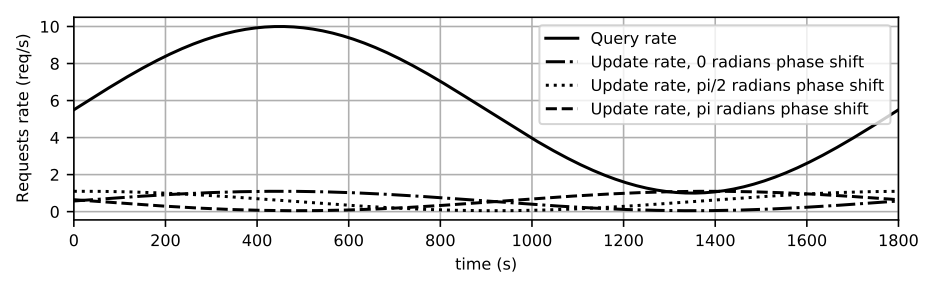}
  \caption{Average request rates (requests/sec) of sinusoidal workloads with the update thread phase shifted 0, $\pi/2$, and $\pi$ radians from the query thread to produce different rate relationships between the two threads. Phase shift by $\pi/4$ radians omitted for graph readability.}
  \label{figSinusoidalWorkloads}
\end{figure*}

%% file: figures/network_reduction_vs_error_fraction.tex
\begin{figure}
  \centering
  \includegraphics[width=\columnwidth]{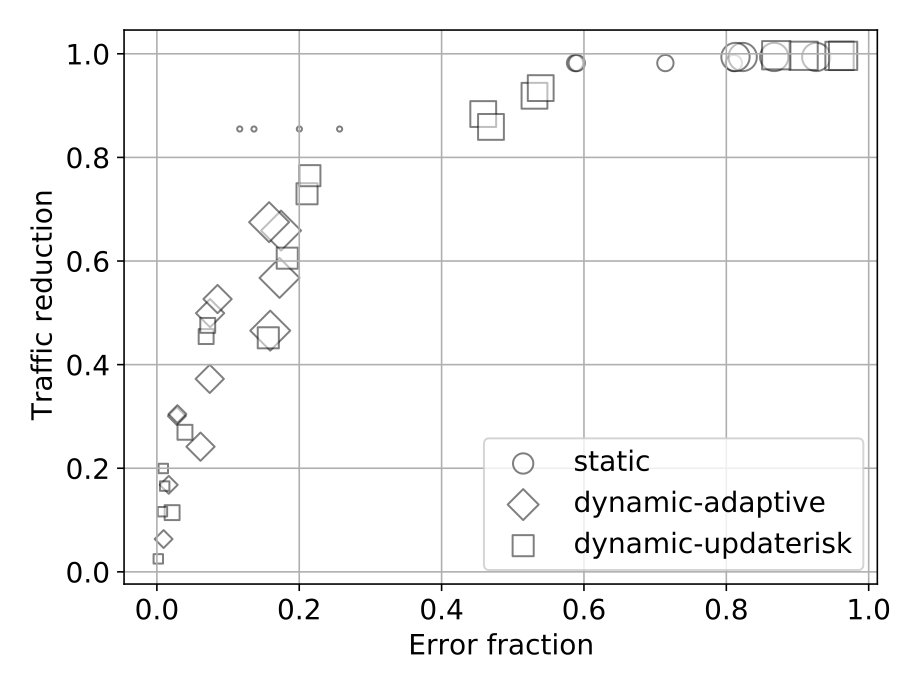}
  \caption{Trade-off between network reduction and error fraction visualized as a scatterplot. All parametrizations and phase shifts are reported per algorithm, and the size of the marker is related to the parameter given to the algorithm. The presented results are averaged over 3 repetitions of each experimental setup.}
  \label{figNetworkReductionVsErrorFraction}
\end{figure}

%% file: figures/dynamic-updaterisk-phaseshift-pi.tex
\begin{figure*}

\centering

\begin{subfigure}[b]{0.48\textwidth}
  \includegraphics[width=\textwidth]{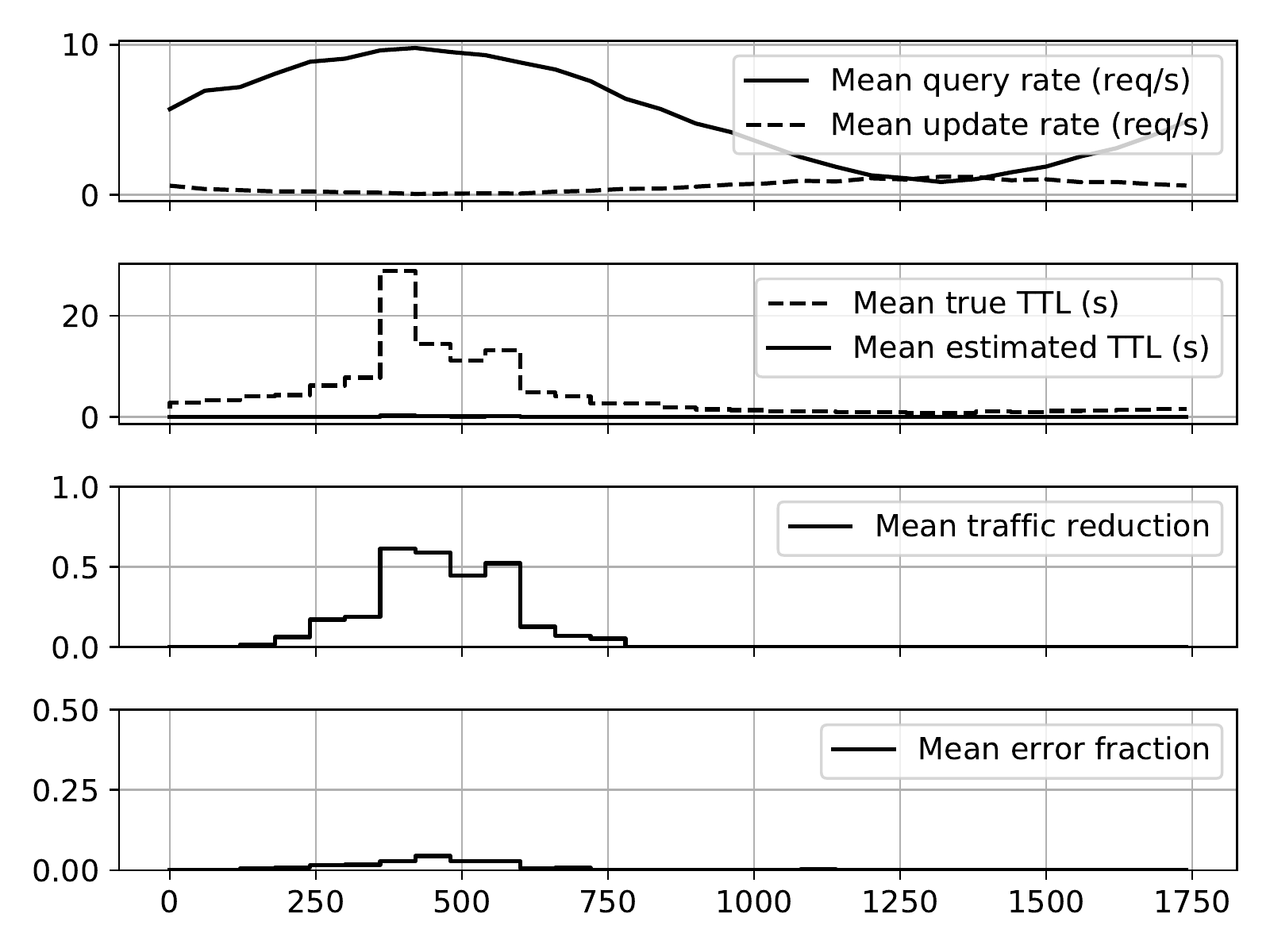}
	\caption{\textbf{dynamic-updaterisk-0.10} ($\rho=0.1$)}
	\label{figUpdateRiskTimeseries01}
\end{subfigure}
~
\begin{subfigure}[b]{0.48\textwidth}
  \includegraphics[width=\textwidth]{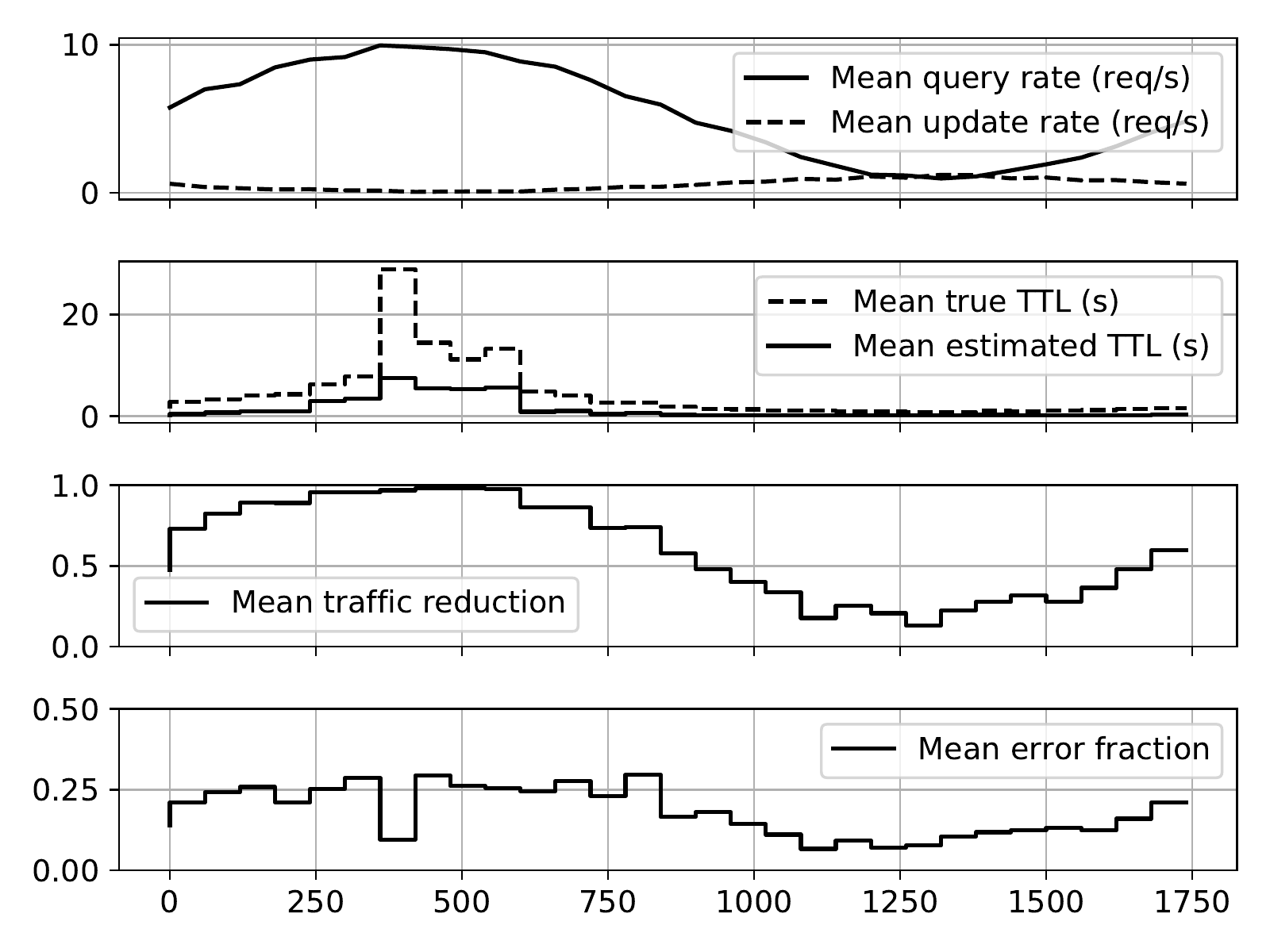}
	\caption{\textbf{dynamic-updaterisk-0.50} ($\rho=0.5$)}
	\label{figUpdateRiskTimeseries05}
\end{subfigure}

\caption{Time series showing behavior of \textbf{dynamic-updaterisk} at two different parametrizations for the same phase shift ($\pi$ in this case), averaged over 3 repetitions.}
\label{figUpdateRiskTimeseries}
\end{figure*}

%% file: figures/hipsterShopOverview.tex
\begin{figure*}
  \centering
  \includegraphics[width=0.85\textwidth]{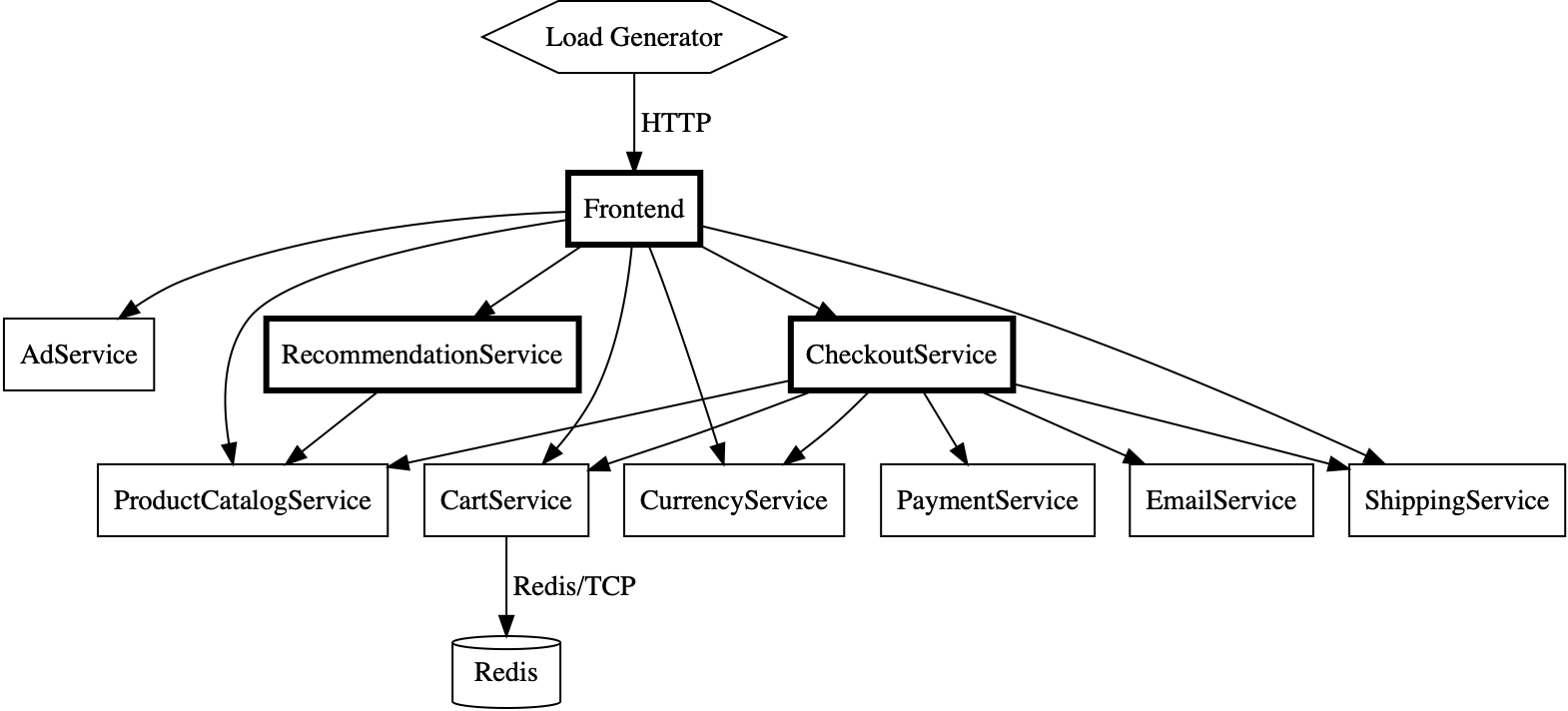}
	\caption{Overview of the Hipster Shop architecture. Unless stated differently, the protocol used for inter-service communication is gRPC. Caches are added to the three components with bold border: Frontend, RecommendationService, and CheckoutService.}
	\label{figHipsterShopOverview}
\end{figure*}

%% file: figures/hipsterShopNetworkTraffic.tex
\begin{figure*}
  \centering
  \includegraphics[width=0.85\textwidth]{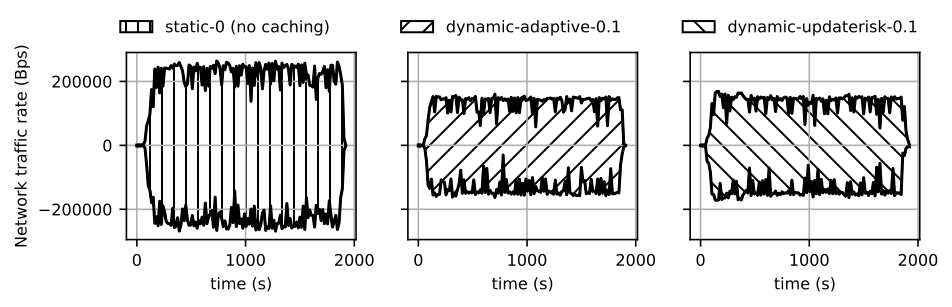}
  \caption{Network traffic over time (bytes/second) in Hipster Shop for when no caching is used (\textbf{static-0}), and for the two most conservative dynamic estimation algorithm parametrizations from Section~\ref{secExperimentsConfigurations} (\textbf{dynamic-adaptive-0.1} and \textbf{dynamic-updaterisk-0.1}). Outgoing traffic is shown as positive, incoming as negative. Thinner overall shape therefore implies less traffic.}
  \label{figHipsterShopNetworkTraffic}
\end{figure*}

%% file: tables/hipsterShopNetworkTraffic.tex
\begin{table*}
\centering
\caption{Inter-Pod network traffic in Hipster Shop. Note that traffic within individual Pods does not count toward these numbers, only traffic between Pods.}
\label{tabHipsterShopNetworkTraffic}
\begin{tabular}{lrrrr}
\toprule
{} & Total bytes & Received bytes &  Sent bytes & Reduction \\
\midrule
\textbf{static-0              } & 54328661.67 &    27382959.33 & 26945701.67 &      0.00 \\
\textbf{dynamic-adaptive-0.1  } & 32335455.67 &    16120919.00 & 16214536.33 &      0.40 \\
\textbf{dynamic-updaterisk-0.1} & 33305797.67 &    16731636.33 & 16574161.00 &      0.39 \\
\bottomrule
\end{tabular}
\end{table*}

%% file: tables/hipsterShopCachedRequests.tex
\begin{table*}
\centering
\caption{Caching of requests in Hipster Shop, as reported by caches installed in the three components that make inter-service requests over gRPC (see Figure~\ref{figHipsterShopOverview}). Note that the total number of requests includes both those that could be cached and those that cannot, see Section~\ref{secHipsterShopCacheable}.}
\label{tabHipsterShopCachedRequests}
\begin{tabular}{lrrrr}
\toprule
{} & Total requests & Cached requests & Upstream requests & Cached request fraction \\
\midrule
\textbf{static-0              } &      254649.67 &            0.00 &         254649.67 &                    0.00 \\
\textbf{dynamic-adaptive-0.1  } &      256923.67 &       207020.33 &          49903.33 &                    0.81 \\
\textbf{dynamic-updaterisk-0.1} &      257745.00 &       203750.33 &          53994.67 &                    0.79 \\
\bottomrule
\end{tabular}
\end{table*}

%% file: sections/relatedwork.tex
\section{Related work}
\label{secRelatedWork}

%

Much of the literature on caching focuses on minimizing storage costs while keeping cache hit ratios at or above a certain level. This is motivated, at least in part, by the fact that efficiencies in this regard directly increase the cost-efficiency of content distribution networks (CDNs). In works such as~\cite{basu2018adaptive,gast_asymptotically_2016,berger_maximizing_2015,berger_exact_2014%
}, the focus is on the cache component and how it prioritizes content such that popular items are more likely to be kept in cache and less popular ones are evicted, even if their server-supplied TTL states that they should still be considered valid. 

In our target domain, we have to assume that the origin server neither knows the true TTL of a response, nor supplies one. 
Lee et al.~\cite{lee2002updaterisk} is the source from which we have implemented both the Adaptive TTL and Update-risk based algorithms. 
The masters thesis by Schaarschmidt~\cite{schaarschmidt2015towards} contains TTL estimation algorithms that make use of machine learning, and are therefore considerably more expensive in terms of both processing and storage than the algorithms selected for our evaluation. 
Fawaz and Artail target a different domain, namely mobile phones relying on cached data during disconnected operation, and thus operate rather as a frontend cache than an inter-service one, they propose an simple exponentially weighted moving average to keep track of update frequency~\cite{fawaz_dcim_2013}. This could very well as future work be added to our caching infrastructure as an option.

In particular, Batchelder et al.\ describe how to determine cacheability~\cite{batchelder_method_2002} and Feiertag et al.\ provide an algorithm for estimating TTL dynamically, based on a hit rate, a change rate, and a ``freshness'' of the data object~\cite{feiertag_updating_2004}. Freshness is partially derived automatically, and partially provided by the developer. As this is likely a reason why inter-service caching is so uncommon, we strongly prefer not placing the burden of the latter on developers or operators.

We view the two caching-related fields as complementary, because smart cache evictions to keep hit rates high and memory use efficient do not make TTL estimates less useful and vice versa. Cache eviction requires some TTL (estimated or accurate) to work with, and estimated TTLs require some kind of cache to be useful. 

Unlike implicit cache maintenance via TTLs, explicit methods rely on the server sending out \emph{invalidation messages} to caches when the underlying data changes. 
This technique improves cache coherency by lowering data staleness~\cite{chenjie_liu_maintaining_1997}. For instance, Cachematic~\cite{holmqvist_cachematic_2019} analyzes SQL queries and infers when changes may have occurred. 
However, multi-hop inter-service communication poses a challenge to the explicit approach, and relies on application-specific knowledge to understand how different operations in different services affect the results of each other. Jia et al.\ leverage deep knowledge into application operation semantics to determine which operations can be cached~\cite{jia_bpcs_2016} to overcome this issue. In contrast, our implicit approach is application-agnostic and needs to neither infer nor be semantically informed about inter-service relationships or functionality.


%% file: sections/summary.tex
\section{Conclusion}
\label{secSummary}

The micro-services paradigm dictates a separation of concerns and strict data ownership, which implies that services must interact frequently with each other across the network. Such communication is increasingly dealt with by service meshes, which uniformly implement features such as load-balancing, retrying, and circuit breaking. Circuit breaking is an effective strategy to conditionally disallow requests toward malfunctioning or overloaded services, thereby increasing overall cloud application resilience.

In this work, we proposed caching as a \emph{soft circuit breaker} actuation mechanism.
We estimate cache TTLs for responses dynamically using adaptive algorithms from the literature on serving web content. We have evaluated our approach and found that in spite of frequent updates, conservative configuration of dynamically estimated TTL estimation algorithms could keep data staleness at 0--3\% while reducing load by up to 30\%. When used in a realistic off-the-shelf e-commerce micro-service application, 80\% of requests were served cached responses and 40\% fewer bytes were transferred.

Completing the soft circuit breaker vision requires an improved decision mechanism, to determine \emph{when} the caching circuit breaker should be activated and by how much. If resources are plentiful, there is no pressing need to introduce possible data staleness errors via caching. In such cases, all requests should be processed using the freshest possible data. But when resources are scarce, a higher risk of data staleness is acceptable. Whether to base such a decision on current resource utilization and thresholds~\cite{larsson_quality-elasticity_2019} or control theory to keep, e.g., response times within a given bound~\cite{klein_brownout_2014}, or via some other mechanism, remains a topic of future work.

\section*{Open Source and Open Data Notice}
\label{secOpenSource}
The source code and data sets used in this work are available in the following locations:

\begin{itemize}
\item \url{https://github.com/llarsson/grpc-caching-interceptors} hosts the gRPC interceptors that implement the Caching and TTL Estimation functionalities.
\item \url{https://github.com/llarsson/protobuf} contains a modified Protobuf compiler that provides a reverse proxy server.
\item \url{https://github.com/llarsson/value-service} hosts the Client and Server of the Value Service. Follow links given therein to the Cache and Estimator repositories.
\item \url{https://github.com/llarsson/value-service-experiments} contains the data set from first suite of experiments.
\item \url{https://github.com/llarsson/hipster-shop} is the Hipster Shop application and our scripts and Kubernetes manifests for running experiments.
\item \url{https://github.com/llarsson/hipster-shop-experiments} contains the data set from the second suite of experiments.
\end{itemize}